\documentclass[aps,showpacs,showeqnos,toolkits,nofootinbib]{revtex4}
\usepackage{mathrsfs}
\usepackage{amsmath,amssymb}
\usepackage{graphicx}
\usepackage{color}
\usepackage{epsfig}
\usepackage{subfigure}
\usepackage{graphics}

\usepackage[usenames]{pstcol}

\def\bea{\begin{eqnarray}}
\def\eea{\end{eqnarray}}
\def\beq{\begin{equation}}
\def\eeq{\end{equation}}

\def\Tr{{\,\mbox{Tr}}}

\def\tr{\mathrm{Tr}}
\def\mui{\mu_I}

\begin{document}

\title{ On Skyrmion semiclassical quantization in the
presence of an isospin chemical potential}

\author{Thomas D. Cohen}
\email{cohen@physics.umd.edu}

\affiliation{Department of Physics, University of Maryland,
College Park, MD 20742-4111, USA}

\author{Juan A. Ponciano}
\email{ponciano@fisica.unlp.edu.ar}

\affiliation{CEFIMAS, Av. Santa Fe 1145, (1059) Buenos Aires,
Argentina\\ CONICET, Rivadavia 1917,
(1033) Buenos Aires, Argentina\\Universidad Nacional de La Plata, C.C. 67, (1900) La
Plata, Argentina}

\author{Norberto N. Scoccola}
\email{scoccola@tandar.cnea.gov.ar}

\affiliation{Physics Depart., Comisi\'on Nacional de Energ\'\i a
At\'omica, (1429) Buenos Aires, Argentina\\ CONICET, Rivadavia 1917,
(1033) Buenos Aires, Argentina\\ Universidad Favaloro, Sol{\'{\i}}s 453,
(1078) Buenos Aires, Argentina}

\begin{abstract}
The semiclassical description of Skyrmions at small isospin
chemical potential $\mu_I$ is carefully analyzed. We show that
when the calculation of the energy of a nucleon is performed using
the straightforward generalization of the vacuum sector techniques
($\mu_I=0$), together with the ``natural" assumption $\mu_I =
{\cal O} (N_c^0)$, the proton and neutron masses are nonlinear in
$\mu_I$ in the regime $|\mu_I| < m_\pi$. Although these
nonlinearities turn out to be numerically quite small, such a result
fails to strictly agree with the very robust prediction that for
those values of $\mui$ the energy excitations above the vacuum are
linear in $\mu_I$. The resolution of this paradox is achieved by
studying the realization of the large $N_c$ limit of $QCD$ in the
Skyrme model at finite $\mui$. This is done in a simplified
context devoid of the technical complications present in the
Skyrme model but which fully displays the general scaling behavior
with $N_c$. The analysis shows that the paradoxical result appears
as a symptom of using the semi-classical approach beyond its
regime of validity and that, at a formal level, the standard
methods for dealing with the Skyrme model are only strictly
justified for states of high isospin $I \sim N_c$.
\end{abstract}


\maketitle

\section{Introduction}

This paper deals with a paradox  associated with the description
of Skyrmions\cite{Skyrme:1961vq,Skyrme:1962vh,Zahed:1986qz} in
systems with a non-zero isospin chemical potential $\mu_I \neq
0$---{\it i.e.}, a chemical potential coupled to the third
component of the isospin\cite{Loewe:2004ic,Ponciano:2007jm}.
Before discussing this in detail it is probably of some utility to
discuss why one should care about the problem. One reason is
simple: the problem of strong interactions with $\mu_I \neq 0$ is
an interesting problem in theoretical physics in its own right,
regardless of whether it has any immediate experimental
implications. From the perspective of QCD there are certain
aspects of the problem which are tractable at both small isospin
chemical potentials and others which are tractable at large
isospin chemical potentials\cite{Son:2000xc,Birse:2001}. Thus the
study of this problem provides an additional avenue to gain
insight into QCD dynamics. From the perspective of the Skyrme
model---whose purpose is to mimic certain key aspects of QCD---the
problem raises a distinct set of theoretical and mathematical
issues which are interesting and illuminating in their own right.
However, there is a more programmatic reason why this problem is
of potential importance. One of the central problems in modern
theoretical nuclear physics is the behavior of hadronic matter in
extreme conditions of temperature or density.  Ultimately,
properties of such matter follows from the underlying theory of
strong interactions, namely QCD.  However, while {\it ab initio}
calculations of QCD at high temperature and zero density are
tractable via lattice simulations, the problem of dense matter is
more problematic as they are afflicted by the notorious fermion
sign problem. There has been considerable recent activity in
attempting to develop viable lattice methods for systems at high
temperature and low but nonzero chemical potential\cite{cp}.
Unfortunately, these methods break down as the temperature
decreases and are unsuitable for the study of cold dense matter.
This is unfortunate since this regime is of potential importance
in astrophysical applications.

Given this situation it is natural to consider models which (it is
hoped) capture much of the essence of QCD, while being tractable
in the regime of interest. One class of models which may be
employed are Skyrme-type models. These have the virtue of encoding
the scaling properties of QCD with $N_c$ (the number of colors).
Moreover, the models become essentially classical at large $N_c$
and thus are far more tractable than fully quantum theories like
QCD.  In fact, over the years there has been a rather large
literature dealing with cold dense Skyrmion matter\cite{sm}

Unfortunately, there is a fundamental difficulty in using the
Skyrme model---or any other model---in place of QCD for cold dense
matter. Even if a model is known to work well in reproducing the
results of nature---and hence of QCD---in the vacuum sector, one
is never sure whether the model is likely to continue to mimic QCD
to good approximation in a quite different regime.  Ideally one
would calibrate the models against QCD in the regime of interest
and test how they do.  However, for cold dense matter this is not
possible; the crux of the issue is that QCD is not tractable and
observational data from astrophysics is, at best, both incomplete
and quite indirect.

In these circumstances it becomes exceptionally important to test
models against QCD anyplace where it is both computationally
tractable and shares features in common with the regime of
interest---cold dense matter.  In this regard QCD with $\mu_I \neq
0$ can play a critical role. On the one hand, unlike the case of a
non-vanishing baryon chemical potential, the functional
determinant in a path integral can be shown to be real and
non-negative for systems with a non-vanishing isospin chemical
potential\cite{Alford:1998sd}.  This means that the fermion sign problem is
evaded and lattice simulations become practical.  Preliminary
simulations have already been done\cite{Kogut:2002zg}.
These were relatively small but increasing accurate lattice studies are possible
\cite{recent}.  On the other hand, this system has {\it something} in
common with the problem of interest---namely a treatment of QCD
with a baryon chemical potential. Thus, studies of models of QCD
in the regime of $\mu_I \neq 0$ can play an important role in
testing models against QCD.

Of course, if one intends ultimately to use a regime with  $\mu_I
\neq 0$ to test a model against QCD, it is essential that one has
computed the  $\mu_I \neq 0$ regime correctly in the model.  It is
in this context we wish to point out a paradoxical result that
arises in the semiclassical treatment of the Skyrme model in the
presence of a finite $\mu_I$.

The paradox concerns the behavior at small but non-zero values of
$\mu_I$ and zero temperature.  At zero temperature the behavior of
QCD is known for small $\mu_I$.  When $| \mu_I | < m_\pi$ the QCD
vacuum state is unaltered by the presence of an isospin chemical
potential.  The {\it only} effect that adding a chemical potential
has in that regime is to shift excitation energies above the
vacuum ({\it e.g.}, single particle energies for hadrons) by
$\mu_I I_3$ where $I_3$ is the isospin of the excited state.  Thus
in this regime all energies are strictly linear in $\mu_I$ with a
slope given by $I_3$.  At $T=0$, nonlinearities only appear for $|
\mu_I | > m_\pi$.  When $ |\mu_I |$ reaches $m_\pi$, the energy
for the lowest pionic excitation reaches zero. Beyond this point
it is energetic favorable for pions to condense. This pion
condensation alters the vacuum structure and thereby alters
excitation energies beyond the linear shift due to the direct
$\mu_I I_3$ contribution.

The paradox is that calculations of the energy of a proton or
neutron in a Skyrme model using the natural generalization of the
techniques used in the vacuum sector yield  proton and neutron
masses which are {\it nonlinear} in $\mu_I$ in the regime $\mu_I <
m_\pi$.  As we will show in Sec. III these nonlinearities are
numerically quite small; however, the linear behavior in $\mu_I$
is a fundamental property of the quantum system the model is
supposed to describe. Thus even small non-linearities appear to
imply that there is something fundamentally wrong with the
approach---or at the very least, something fundamental which is
not fully understood.

This is quite problematic on two grounds.  The first is what was
alluded to above.  In order to use the $\mu_I \neq 0$ regime as a
test of Skyrme type models in regimes where chemical potentials
play a role, it is essential to have reliable calculations. The
second is more vexing: if the standard approach fails at some deep
level in its straightforward generalization from the vacuum sector
to $\mu_I \neq 0$, how certain can one be of its validity in the
vacuum sector?

In this paper we resolve this paradox---at least at a formal
level.  The implication of this resolution for phenomenology,
however, remains somewhat unclear.

The crux of the resolution involves the nature of the large $N_c$
limit of QCD.  The standard methods for dealing with the Skyrme
model---finding a classical static soliton solution and then
requantizing the collective modes to restore symmetries and
describe physical states---are only strictly justified at large
$N_c$.  Thus in using these methods one is implicitly studying a
large $N_c$ world and then attempting to identify features which
extrapolate back to the real world of $N_c=3$.  One feature of
this large $N_c$ world is in an infinite tower of baryon states
with $I=J$.  For $N_c$ large but finite, the levels in this tower
are split by effects of order $1/N_c$.  However these effects grow
with $I$ (or equivalently $J$).  For sufficiently large  $I$, the
excitation energies become too large to neglect and the
semiclassical methods based on the $1/N_c$ expansion ultimately
break down. The usual interpretation of the $I=J$ tower of states
for the physical world of $N_c=3$ is that the $I=J=1/2$ states
correspond to the nucleon, the $I=J=3/2$ states to the delta,
while states with $I=J>3/2$ correspond to the region beyond the
scope of the approximation and are artifacts of the large $N_c$
limit.

In considering the formal large $N_c$ limit, it is important to
recall that there are critical ways in which the large $N_c$ limit
and the chiral limit do not commute\cite{TDC}.  The standard
semiclassical treatment of the Skyrme model is based on a pure
$1/N_c$ expansion.  Thus implicitly when studying chiral
properties in Skyrme models using standard quantization methods
one is implicitly taking the large $N_c$ limit before the chiral limit.
Thus at least formally, $m_\pi \sim N_c^0$, even though $m_\pi$ is
numerically small. This formal fact plays an essential role in
understanding the origin of non-linearities for $|\mu_I| < m_\pi$.

The resolution to the paradox at the formal level is the
following: when $\mu_I$ is treated as being of order $N_c^0$ (but
less than $m_\pi$), the natural generalization of the methods used
in the vacuum {\it is} formally correct at large $N_c$ for the
state of the lowest free energy $E - \mu_I I$ and for a band of
low-lying states above it.  However, these states are formally
states of high isospin---$I \sim N_c$---and as such do not include
the nucleon. Thus in a formal sense (based on $1/N_c$ expansion),
the nucleon is outside the domain of validity of the standard
semiclassical methods. The non-linearities in $\mu_I$ are a
symptom of using the semiclassical approach beyond it regime of
validity.

The plan of this paper is as follows:  The next section contains a
brief discussion of why the hadrons' energies are linear in $
\mu_I $ for $| \mu_I | < m_\pi$.  The following section contains
a review of the semiclassical treatment of the nucleon in the
Skyrme model; this yields nonlinearities in $\mu_I$  for $| \mu_I|
< m_\pi$, provided $\mu_I$ is treated as being formally of order
$N_c^0$. This is in contrast to the general results of the
preceding section and as such constitutes the heart of the
apparent paradox. In the section following, a simple toy quantum
mechanical model is introduced.  This model has central features
analogous to the soliton and its semiclassical treatment but in a
far more transparent form; it illustrates explicitly the
underlying issues. The model has the virtue of being tractable
quantum mechanically as well as semiclassically and thus enables
to explicitly verify one's understanding.  These model results
illustrate the resolution of the paradox mentioned in this
introduction. A final section discusses the implication of these
results.

\section{The quantum mechanics of a system in the presence of
an isospin chemical potential\label{formal}}

In analyzing the Skyrme model at finite isospin chemical potential
it is essential to recall that although the model is typically
treated classically it is ultimately being used to model a quantum
system---namely QCD.  Accordingly it is essential to ensure that
any treatments of the model correctly encode the underlying
quantum mechanics.  In this section, we review why elementary
quantum considerations require that energy of the proton and the
neutron are necessarily linear in $\mu_I$ below a critical value
(provided that $T=0$, as will be assumed throughout this paper).

To begin let us consider the problem a bit abstractly.  Let us
consider a general quantum mechanical system with Hamiltonian,
$\hat{H}_0$. The system can be quite general with one important
proviso---there must be a gap in the spectrum between the ground
state and the lowest excited states. Suppose further that there is
some operator, $\hat{Q}$ which corresponds to a conserved
``charge''.
\begin{equation}
[\hat{H}_0,\hat{Q}]=0 \,.
\end{equation}
For the present problem $\hat{Q}$ is the third component of the
isospin (which is a conserved quantity in both QCD and  in the
Skyrme model). Because $\hat{H}_0$ and $\hat{Q}$ commute they have
simultaneous eigenstates (if there are degeneracies in the
spectrum of $\hat{H}_0$ we will work on a basis that where
$\hat{H}_0$ and $\hat{Q}$ commute they have simultaneous
eigenstates):
\begin{equation}
\hat{H}_0 |n, q \rangle = E_{n,q} |n, q \rangle \qquad , \qquad
\hat{Q} |n,q \rangle = q |n, q \rangle, \label{eigen}
\end{equation}
where $q$ labels the charge of the state and $n$ labels all other
quantum numbers.

Adding a chemical potential for $\hat{Q}$ to the Hamiltonian
yields a new Hamiltonian
\begin{equation}
\hat{H}'=\hat{H}_0 - \mu_Q \hat{Q} \,.
\end{equation}
This new Hamiltonian is a free energy operator; its ground state
corresponds to the free energy of a system at fixed chemical
potential and zero temperature.  For field theories such as QCD
(or the Skyrme model) it is the vacuum in the presence of the
chemical potential. The commutativity of $\hat{H}_0$ with $\hat{Q}$
ensures that the eigenstates of $\hat{H}_0$ are also eigenstates
of $\hat{H}'$.  Equation (\ref{eigen}) implies that its
eigenvalues are given by
 \begin{equation}
\hat{H}' |n, q \rangle = E'_{n,q} |n, q \rangle \; \;\ {\rm with}
\; \; E'_{n,q}=(E_{n,q}-\mu_Q).
 \label{eigen2}\end{equation}

Equation (\ref{eigen2}) immediately implies that the absolute energy
of {\it any state} depends linearly on $\mu_Q$ for any value of
$\mu_I$. It is important to note that  this does {\it not} imply
that {\it excitation energies} are also linear in $\mu_Q$ for any
value $\mu_Q$.  The reason for this is that excitation energies
are measured relative to the ground state energy and ground state
energies need not be linear in $\mu_Q$.  Nonlinear behavior in the
ground state energy occurs due to level crossing: the state which
minimizes $\hat{H}'$ at one $\mu_Q$ need not minimize it another.
In the context of quantum field theories this corresponds to a
phase transition to a new ``vacuum'' state with a condensate
carrying the charge $Q$. However, provided that in a regime in
which the ground state is unchanged from $\mu_Q=0$, the excitation
energies are linear in $\mu_Q$.  For theories with a gap in the
spectrum one expects a finite region in  $\mu_Q$ over which the
vacuum is unchanged and, accordingly, excitation energies are
linear.
Now let us return to the issue of QCD  with an isospin chemical potential and zero temperature.
The proton and neutron are excitations with baryon number of unity and a third component of
isospin of $\pm 1/2$ above the vacuum.  Provided one is below the critical isospin chemical potential
for a phase transition, the previous argument ensures the excitation energies for these must be linear
in $\mu_I$.  The critical issue is the value of $\mu_I$ for which the phase transition occurs. There is
very strong evidence that critical isospin chemical potential occurs at $\mu_I=\pm m_\pi$; for $|\mu_I|$
charged pions condenses and the vacuum structure alters. This is precisely the picture seen in  vacuum models
based on chiral perturbation theory\cite{Son:2000xc}.  Thus, as a very robust prediction one expects the energy
of the proton and neutron excitation above the vacuum to be linear in $\mu_I$ for $|\mu_I| < m_\pi$.

One might worry that there is no rigorous demonstration that the
critical chemical potential is at $|\mu_I|=  m_\pi$.  In order for
the critical chemical potential to be smaller than this, there
must be a state in QCD with mass per unit isospin {\it less} than
$m_\pi$; this is highly implausible on its face.  Moreover even if
this highly implausible scenario were true, it would not alter the
formal problem with the Skyrme model calculations.  In those model
calculations the vacuum does not undergo a phase
transition---vacuum properties far from the soliton are unaltered
by the chemical potential.  Thus for the model the proton and
neutron must be linear in $\mu_I$ and yet, as we will see in the
following section, { a naive extension of the standard
semiclassical treatment lead to deviations from linearity.}

\section{Semiclassical treatment of the Skyrme model at $\mu_I \neq 0$}

In this section we analyze the semiclassical quantization of the
skyrmion paying special attention to the issues raised by the presence
of a finite $|\mui| \leq m_\pi$. For definiteness we consider here the
lagrangian of the $SU(2)$ Skyrme model with quartic term stabilization
and finite pion mass. It is given by
\begin{eqnarray}
{\cal L} = - \frac{f_\pi^2}{4} \tr \left\{ L_\alpha L^\alpha
\right\} + \frac{1}{32e^2} \tr \left\{ [L_\alpha,L_\beta]^2
\right\} + \frac{m_\pi^2 \ f_\pi^2}{4} \tr\left\{ U + U^\dagger -
2 \right\}. \label{skyr}
\end{eqnarray}
In Eq.(\ref{skyr}), as usual, $U$ represents the
$SU(2)$ chiral field and the Maurier-Cartan operator $L_\alpha$ is
defined by $L_\alpha = U^\dagger
\partial_\alpha U$.
The isospin chemical potential $\mui$ is introduced by performing
the replacement
\begin{equation}
\partial_\alpha U \longrightarrow \partial_\alpha U -
i \ \frac{\mui}{2} \ [\tau_3 , U] \ g_{\alpha 0},
\end{equation}
where $g_{\alpha\beta}$ is the metric tensor in Minkowski space
and $\tau_3$ is the third Pauli matrix. In what follows we will
assume that $|\mui| < m_\pi$, namely that the perturbative vacuum
$U=1$ is stable against pion condensation.

We consider a spinning soliton configuration of the form
\begin{equation}
U = A \ \tilde U(R^{-1} \vec r) \ A^\dagger ,
\end{equation}
where $A$ and $R$ are time-dependent isospin and spin rotations,
respectively. $\tilde U (\vec r)$ is a static configuration.
Inserting this form for $U$ in Eq.(\ref{skyr}) we get
\begin{equation}
L = - M_0 +
\frac12 \Lambda^I_{ab} \left( \omega_a - \mui D_{3a} \right) \left( \omega_b - \mui D_{3b} \right)
+
\frac12 \Lambda^J_{ij} \ \Omega_i \ \Omega_j
+ \Lambda^M_{ai} \  \left( \omega_a - \mui D_{3a} \right) \ \Omega_i .
\label{ll}
\end{equation}
Here, we have used
\begin{eqnarray}
A^{-1} \ \dot{A} = \frac{i}{2} \ \omega_a \ \tau_a , \nonumber \\
\left( R^{-1} \ \dot{R} \right)_{ij} = \epsilon_{ijk} \ \Omega_k ,
\end{eqnarray}
and
\begin{eqnarray}
D_{ab} = \frac12 \tr \left[ \tau_a \ A \tau_b \ A^{-1} \right].
\end{eqnarray}
The static mass $M_0$ is given by
\begin{eqnarray}
M_0 &=& - \int d^3r \left[ \frac{f_\pi^2}{4} \tr \left\{ \tilde L_i \tilde L_i
\right\} + \frac{1}{32e^2} \tr \left\{ [\tilde L_i,\tilde L_j]^2 \right\} +
\frac{m_\pi^2 \ f_\pi^2}{4} \tr\left\{\tilde U + \tilde U^\dagger - 2
\right\}\right],
\end{eqnarray}
while the tensors $\Lambda^I_{ab}$, $\Lambda^J_{ab}$ and
$\Lambda^I_{ab}$ are the isospin, spin and mixed inertia tensors,
respectively, given by
\begin{eqnarray}
\Lambda^I_{ab}&=&\int d^3r \ \left[\frac{f_\pi^2}{8}
 \Tr\left\lbrace\tilde U^\dagger[\tau_a, \tilde U] \tilde U^\dagger[\tau_b, \tilde U]\right\rbrace+
 \frac{1}{32e^2}\Tr\left\lbrace \left[ \tilde L_i,\tilde U^{\dagger}[\tau_a, \tilde U]\right] \left[ \tilde L_i,\tilde U^{\dagger}[\tau_b, \tilde U]\right]  \right\rbrace \right], \\
\Lambda^J_{kl}&=&\int d^3r \ \epsilon_{ijk}
\epsilon_{mnl}\ r_j \ r_n \left[
 -\frac{f_\pi^2}{×2}\tr\left\lbrace \tilde L_i \tilde L_m\right\rbrace -\frac{1}{8e^2}\Tr\left\lbrace [\tilde L_i, \tilde L_a][\tilde L_m,\tilde L_a]\right\rbrace \right], \\
 \Lambda^M_{ka}&=&\int d^3r \ i\epsilon_{ijk} \ r_j \left[   \frac{f_\pi^2}{4}\Tr\left\lbrace \tilde U^{\dagger} [\tilde U,\tau_a] \tilde L_i\right\rbrace +
 \frac{1}{16e^2}\Tr \left\lbrace \left[ \tilde U^\dagger [\tilde U,\tau_a],\tilde L_m\right] [\tilde L_i, \tilde L_m]\right\rbrace
\right].
\end{eqnarray}
The number of independent non-vanishing components of the inertia
tensors is given by the explicit form of the static ansatz. All
the configurations to be considered below will be symmetric under
$\pi$-rotations along any of the three cartesian axes. In this
case one can prove that all the inertia tensors are diagonal.
Thus in what follows we will assume that
\begin{equation}
\Lambda^I_{ab} = \Lambda^J_{ab} = \Lambda^M_{ab} =0 \qquad
\mbox{if}\ a \neq b.
\label{ndiad}
\end{equation}

Now let us pay attention to the $N_c$-order of the different quantities
appearing in Eq.({\ref{ll}). As usual, $M_0$ and the inertia tensors
$\Lambda$'s are of ${\cal O}(N_c)$ while $\omega_a$ and $\Omega_i$ are
taken to be of ${\cal O}(N_c^{-1})$.
In order to proceed we have to determine which order in
$N_c$ has to be assigned to $\mui$. Since we are interested
in values of $\mui \leq m_\pi$ it appears to be natural to
take $\mui = {\cal O}(N_c^0)$. In this case, to leading order in $N_c$
Eq.({\ref{ll}) reads
\begin{equation}
L^{(1)} = - M_0 + \frac12 \ \mui^2 \sum_{a=1,2,3}\ \Lambda^I_{aa} \ D_{3a} D_{3a},
\label{l1}
\end{equation}
where Eq.(\ref{ndiad}) has been used. In principle the equations
to determine the static soliton configuration should be obtained
by minimizing this lagrangian. Some simplification can be
performed by noting that since the chemical potential acts along
the 3-axis in isospin space, the resulting configuration is
expected to be axially symmetric along the third axis. In this
case we have
\begin{eqnarray}
\Lambda^I_{33} &=& \Lambda^J_{33} = \Lambda^M_{33} = \Lambda_3,  \nonumber\\
\Lambda^I_{11} &=& \Lambda^I_{22} = \Lambda_I,  \nonumber \\
\Lambda^J_{11} &=& \Lambda^J_{22} = \Lambda_J,  \nonumber \\
\Lambda^M_{11} &=& \Lambda^M_{22} = \Lambda_M. \label{axial}
\end{eqnarray}
Using $\sum_{a=1,2,3} D_{3a} D_{3a} = 1$, it is not difficult to
find that Eq.(\ref{l1}) can be expressed as
\begin{equation}
L^{(1)} =  - M_0 + \frac12 \   \Lambda_3 \ \mui^2 + \frac12  \ (\Lambda_3 - \Lambda_I) \ \mui^2 \ \sin^2 \beta .
\end{equation}
Here, we have used the standard parametrization of the Wigner
D-functions in terms of the Euler angles $\alpha,\beta,\gamma$. As
we see, static mass $M_{st} = - L^{(1)}$ depends on the
orientation of the soliton in flavor space. However, if we assume
that the deformation induced by the chemical potential is not too
large we have that $\Lambda_3 - \Lambda_I << \Lambda_3$. In this
case the $\beta$-dependent term can be neglected and we have
\begin{equation}
M_{st} = M_0 - \frac12 \  \Lambda_3  \ \mui^2  .
\label{static}
\end{equation}
Namely, at the static level we have to minimize a $\mui$-dependent
mass which, obviously, leads to a $\mui$-dependent soliton configuration.
In what follows we will denote with an extra upper index the
soliton quantities calculated with this $\mui$-dependent soliton
configuration. Thus, the full lagrangian up to order $1/N_c$ reads
\begin{eqnarray}
L &=& - M_0^{(\mui)} + \frac12 \ \Lambda_I^{(\mui)}\left[ \left( \omega_1 - \mui \ D_{31} \right)^2 +
                                              \left( \omega_2 - \mui \ D_{32} \right)^2 \right]
                    + \frac12 \ \Lambda_J^{(\mui)} \left[ \Omega_1^2 + \Omega_2^2 \right]
                    + \frac12 \ \Lambda_3^{(\mui)}  \left( \omega_3 - \mui \ D_{33} + \Omega_3\right)^2 \nonumber
             \\
 & & \qquad  \qquad  \qquad  \qquad + \Lambda_M^{(\mui)} \left[ \left( \omega_1 - \mui \ D_{31}\right) \ \Omega_1 +
                                              \left( \omega_2 - \mui \ D_{32} \right)\ \Omega_2 \right].
\end{eqnarray}
Defining the canonical conjugate momenta in the usual way, we have
\begin{eqnarray}
I^{bf}_{a} &=& \frac{\partial L}{\partial \omega_{a}} = \Lambda_I^{(\mui)} \left( \omega_a - \mui \ D_{3a} \right) +
                        \Lambda_M^{(\mui)} \ \Omega_a \qquad ; \qquad \mbox{for} \qquad a=1,2 \label{mom1}\\
J^{bf}_{a} &=& \frac{\partial L}{\partial \Omega_{a}} = \Lambda_M^{(\mui)} \left( \omega_a - \mui \ D_{3a} \right) +
                        \Lambda_J^{(\mui)} \ \Omega_a \qquad ; \qquad \mbox{for} \qquad a=1,2 \label{mom2}\\
I^{bf}_3 &=& J^{bf}_3 = \frac{\partial L}{\partial \omega_{a}} =  \frac{\partial L}{\partial \Omega_{a}} =
                   \Lambda_3^{(\mui)} \left( \omega_3 - \mui \ D_{33} + \Omega_3 \right). \label{mom3}
\end{eqnarray}

Here, the upper index $bf$ indicates that these momenta are
defined in the body-fixed frame. Thus, the corresponding
hamiltonian $H=\sum_{a=1,2,3} \left( \omega_a \ I^{bf}_a +
\Omega_a \ J^{bf}_a \right) - L$ reads
\begin{eqnarray}
H &=& M_0^{(\mui)} + \mui \ \sum_{a=1,2,3} D_{3a} J_a^{bf}  \nonumber \\
  & & + \frac{1}{2} \left[ \frac{ \left(J_3^{bf}\right)^2}{\Lambda_3^{(\mui)}} + \frac{
                                   \Lambda_I^{(\mui)} \left[\left(J_1^{bf}\right)^2 + \left(J_2^{bf}\right)^2\right] +
                                    \Lambda_J^{(\mui)} \left[\left(I_1^{bf}\right)^2 + \left(I_2^{bf}\right)^2\right]-
                                     2 \ \Lambda_M^{(\mui)} \left[J_1^{bf} I_1^{bf} + I_2^{bf} J_2^{bf}\right]}
                                     { \Lambda_I^{(\mui)} \Lambda_J^{(\mui)} -  \left( \Lambda_M^{(\mui)} \right)^2 }
                                     \right].
\end{eqnarray}
It should be stressed that, as expected, using
Eqs.(\ref{mom1})-(\ref{mom3}) one can easily check that to leading
order in $N_c$ the minimization of $H$ leads to the same result as
the minimization of $L^{(1)}$. Defining $\vec T = \vec I - \vec J$
we obtain that $H$ can be expressed as
\begin{eqnarray}
H &=& M_0^{(\mui)} - \mui \ I_3 +
  \frac{1}{2} \left[ \frac{ \left( \Lambda_I^{(\mui)} -  \Lambda_M^{(\mui)} \right)  J^2  +
                                 \left( \Lambda_J^{(\mui)} -  \Lambda_M^{(\mui)} \right)  I^2 }
                                 {\Lambda_I^{(\mui)} \Lambda_J^{(\mui)} -  \left( \Lambda_M^{(\mui)} \right)^2 } +
                                 \left( \frac{1}{\Lambda_3^{(\mui)}} + \frac{ 2 \Lambda_M^{(\mui)} -
                                                         \Lambda_I^{(\mui)} - \Lambda_J^{(\mui)} }
   {\Lambda_I^{(\mui)} \Lambda_J^{(\mui)} -  \left( \Lambda_M^{(\mui)} \right)^2 } \right) \left(J_3^{bf}\right)^2 \right]
\nonumber \\
  & & \qquad \qquad \qquad \qquad \qquad + \frac{1}{2} \frac{ \Lambda_M^{(\mui)} }
                                 {\Lambda_I^{(\mui)} \Lambda_J^{(\mui)} -  \left( \Lambda_M^{(\mui)} \right)^2 } \ T^2,
\end{eqnarray}
where we have used the well-known relation $I_3 = - \sum_{a=1,2,3}
D_{3a} J_a^{bf}$ between the lab-frame components of the isospin
and those of the spin in the body-fixed frame. It should be noted
that, of course, $|I-J| \leq T \leq I + J$. This means that for
the particular case of the nucleon ($J=I=1/2$) we have $T=0,1$.
Since for not-too-deformed configurations $\Lambda_I^{(\mui)}
\approx \Lambda_J^{(\mui)} \approx \Lambda_M^{(\mui)}$, we expect
that the mass of the $T=1$ state will be much higher than that of
the $T=0$. Thus, for the physical nucleon we take $T=0$. In this
case, using $J=I=1/2, \left(J_3^{bf}\right)^2 =1/4$ we obtain that
\begin{equation}
M_{(\begin{array}{c}
  {\scriptstyle p} \\
  {\scriptstyle n} \\
\end{array})} =  M_0^{(\mui)} \mp \frac{\mui}{2} + \frac{1}{4}
\left( \frac{1}{2 \Lambda_3^{(\mui)}} - \frac{ 2 \Lambda_M^{(\mui)} -
                                                         \Lambda_I^{(\mui)} - \Lambda_J^{(\mui)} }
           {\Lambda_I^{(\mui)} \Lambda_J^{(\mui)} -  \left( \Lambda_M^{(\mui)} \right)^2 }
                                                         \right).
\end{equation}

So far we have considered the axial symmetric soliton
configuration which is expected to be the lowest energy
configuration in this case. As an approximation to this, in
Ref.\cite{Ponciano:2007jm} a spherical symmetric hedgehog
configuration was used. For completeness, we provide here the
corresponding formulae. In such case, we have
\begin{equation}
\Lambda^I_{ij} = \Lambda^J_{ij} = \Lambda^M_{ij} = \Lambda_0 \ \delta_{ij}.
\end{equation}
Thus, following the same steps as before we get
\begin{equation}
M_{(\begin{array}{c}
  {\scriptstyle p} \\
  {\scriptstyle n} \\
\end{array})} =  M_0^{(\mui)} \mp \frac{\mui}{2} + \frac{3}{8 \Lambda_0^{(\mui)}},
\end{equation}
where $M_0^{(\mui)}$ is related with $M_H$ calculated
in Ref.\cite{Ponciano:2007jm} by
\begin{equation}
M_0^{(\mui)} = M_H(\mui) + \frac12 \ \Lambda_0^{(\mui)} \ \mui^2.
\label{hed}
\end{equation}

The $\mui$ dependence of the proton and neutron
shifts ($\Delta_p$ and $\Delta_n$, respectively)
of the corresponding masses from their values at $\mu_I=0$
is shown in Fig. \ref{nucleonmass}. To perform the numerical calculations
we have used the standard set of parameters of Ref.\cite{ANW}: $f_\pi
= 54\ MeV$; $e= 4.84$; $m_\pi = 138 \ MeV$. We can readily see
that for both the axially symmetry exact configuration (dashed
line) and for the spherical approximate configuration (dotted
line) there is a nonlinear behavior. In fact, the results for
these two configurations are almost identical, with small
differences appearing only for $\mui/m_\pi$ larger than 0.9. This
provides an a posteriori justification for the use of the
spherical approximate configuration made in
Ref.\cite{Ponciano:2007jm}. While the deviation from linearity
(full line) is numerically quite small, the very existence of such
deviation implies an inconsistency with the quantum mechanics
analysis discussed in the previous section. The fact that the
``natural" choice $\mui = {\cal O}(N_c^0)$ leads to such an
inconsistency represents the paradox which we would like to
clarify in this article.

\begin{figure}
\centering
\includegraphics[width=5in]{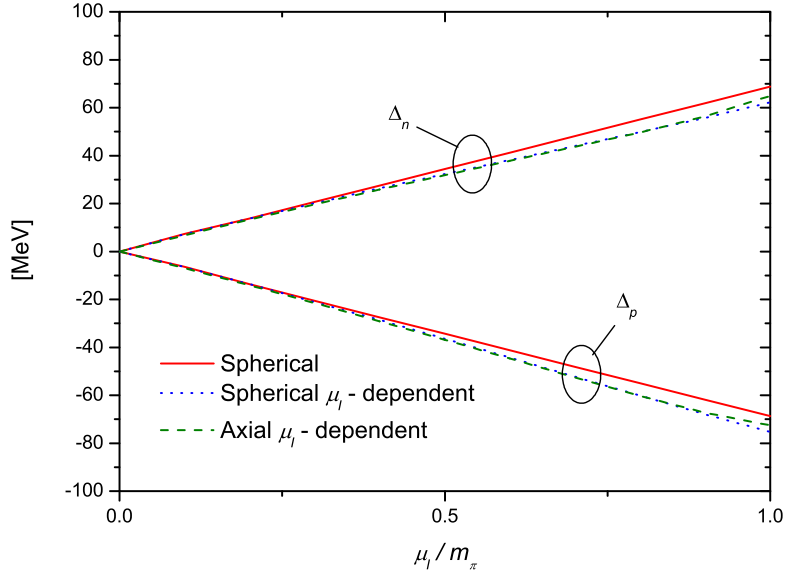}
\caption{Shift of the proton and neutron masses with respect to their corresponding values at $\mu_I=0$ as a function of the
isospin chemical potential $\mu_I$ for $\mu_I \leq m_\pi$. Full line corresponds to the result obtained under the assumption $\mui = {\cal O}(N_c^{-1})$
while the dashed (dotted) line corresponds to the use of axial (spherical) ansatz assuming $\mui = {\cal O}(N_c^{0})$.}
\label{nucleonmass}
\end{figure}

One might wonder whether there is a way to perform the
semiclassical quantization which leads to a linear behavior. In
fact, this is possible provided one makes the somewhat unnatural
assignment $\mui = {\cal O}(N_c^{-1})$. Then, to leading order in
$N_c$, the lagrangian is given by
\begin{eqnarray}
L^{(1)} &=& - M_0
\end{eqnarray}
Therefore, in this case to obtain the static soliton configuration
one should only minimize $M_0$. In this way one obtains a static
configuration $U_{st}$ which is {\it independent} of $\mui$ and,
thus, of the spherically hedgehog type. The full expression for
$L$ reads
\begin{equation}
L = - M_0  + \frac12 \Lambda_0 \ \left( \Omega_a + \omega_a - \mui D_{3a} \right)
\left( \Omega_a + \omega_a - \mui D_{3a} \right)
\end{equation}
where we note that now $M_0$ and $\Lambda_0$ do not depend on $\mui$.
As before we define the momenta
\begin{eqnarray}
I_a \equiv \frac{\partial L}{\partial \omega_a}
=
J_a \equiv \frac{\partial L}{\partial \Omega_a}
= \Lambda_0 \left( \Omega_a + \omega_a - \mui \ D_{3a} \right)
\label{eq15}
\end{eqnarray}
Thus, the resulting Hamiltonian $H = \Omega \ I - L$ is
\begin{eqnarray}
H= M_{sol} +  \mui \ D_{3a} \ J_a + \frac{I_a^2}{2 \Lambda_0}
\end{eqnarray}
which leads to
\begin{eqnarray}
M_{(\begin{array}{c}
  {\scriptstyle p} \\
  {\scriptstyle n} \\
\end{array})} =  M_0 \mp \frac{\mui}{2} + \frac{3}{8 \Lambda_0}.
\end{eqnarray}
Comparing with the situation described above we see that
here there is not intrinsic dependence of $M_0$ and $\Lambda_0$ on $\mui$.
In this case we obtain the energy shift linear in $\mu_I$
represented by a full line in Fig.\ref{nucleonmass}. Although
this might be a resolution of the problem one is still left with the question of why the ``natural"
assignment $\mui = {\cal O}(N_c^{0})$ is not fully compatible with general expectation from general
quantum mechanical principles while the ``unnatural" assignment $\mui = {\cal O}(N_c^{-1})$ does.
In the following section we will address this issue in detail.

\section{A Toy Model}
As seen in the previous section the results from the semiclassical quantization of the Skyrme model with the ``natural" assignment $\mui = {\cal O}(N_c^{0})$ are inconsistent with the formal requirement of the analysis in Sect.~\ref{formal} that the proton and neutron energies are strictly linear in $\mu_I$ below the critical value of $m_\pi$. However, the calculation in the Skyrme model is technically complicated and these complications may tend to obscure the origin of the discrepancy.  These complications are of two sorts.  The first is that the system is a field theory with an infinite number of degrees of freedom.  A few of these are collective modes which are requantized, but the remainder are internal degrees of freedom associated with the shape of the soliton. These internal degrees of freedom come into play when one alters the profile in response to the inclusion of $\mu_I$. The second complication is that collective degrees of freedom---those which are requantized---do not commute with one another so the structure of the quantum mechanical system is nontrivial.

However, the general scaling behavior of the system with
$N_c$---which we believe to be the core of the problem---can be
illustrated in a much simpler context: a quantum system with two
degrees of freedom; one which plays the role of the internal
degrees of freedom and the other the collective ones. This system
is supposed to be analogous to the soliton excitations acting
above the vacuum and hence the analogy only holds in the regime
where the isospin chemical potential has not induced a phase
transition of the vacuum---which is the regime where our paradox
has been identified.   The absolute energies in this model thus
correspond to the excitation energy of the baryons above vacuum.
The collective degree of freedom in the model is associated with a
conserved quantity (which we will take to be the analog of the
third component of isospin); we can add a chemical potential for
this quantity. The system can be chosen so that there is a direct
analog of $N_c$ which controls the extent to which the quantum
mechanical system can be treated semiclassically.  Such a ``toy
model'' is a useful place to test approximation schemes since
{ it is} solvable at the quantum level and, thus, one can
directly compare approximate with exact results.

The model, in units with $\hbar =1$, is given by the following
Hamiltonian which corresponds to a particle of unit charge moving
in two dimensions:
\begin{equation}
\hat{H} = \frac{1}{2 { m}} \left
[\hat{\vec{p}}-\vec{A}(\hat{\vec{r}})\right ]^2+ \lambda V \left (
\lambda^{-1} \, \hat{r}^2 \right )
\end{equation}
where $\lambda$ plays the role of $N_c$ and $\vec{A}$ is an
infinitely thin magnetic flux localized at the origin and with a
strength of precisely { one half of} a flux quantum: $\vec{A}(\vec{x}) =
{ -} \frac{\hat{x} y - \hat{y} x}{2 r^{ 2}} $ where $\hat{x}$ and
$\hat{y}$ indicate unit vectors (as opposed to quantum operators).
The purpose for including this vector potential will become clear
presently. Note that this model is axially symmetric and hence
commutes with $\hat{L}$, the two-dimensional angular momentum; the
Hamiltonian can be written as $\hat{H} = \hat{H}_r +
\frac{{ \left(\hat{L}-1/2\right)^2}}{2 { m r^2}}$ where $\hat{H}_r $ acts entirely on the
radial degrees of freedom  and the $\hat{L}-1/2$ structure
reflects the presence of the vector potential.  It is convenient
to introduce a new operator $\hat{I}\equiv \hat{L} -1/2$.  Note
that a well-defined Hilbert space ({\it i.e.}, a single-valued
wave function) requires that the eigenvalues of $\hat{L}$ be
integers; accordingly the eigenvalues of $\hat{I}$ are $\pm 1/2,
\pm 3/2 , \pm 5/2 \cdots$. Thus $I$ plays the role of $I_3$ in the
Skyrme model and the vector potential was included to ensure
half-integral values for $I$.

Note that the model has a magnetic flux localized at the origin.
Accordingly, the vector potential is undefined at the origin.
Thus, to be sensible our model must exclude the particle from
hitting the origin.  This can be achieved by choosing a form for
$V$ which is repulsive at the origin and sufficiently singular.
Having a singular repulsive potential at the origin has another
virtue: if there is attraction elsewhere in the potential, there
will be a minimum of the potential as a function of $r$.  A
classical solution localized (in $r$) at this minimum plays the
role of the soliton;  angular motion is then the collective
motion.

A convenient choice for the functional form of the potential is
$V(r^2) = \frac{1}{2 m r^2} + \frac{1}{2} m \omega^2 r^2$, since
this system is exactly solvable. Including our scaling factor
$\lambda$ (playing the role of $N_c$) the effective potential for
radial motion associated with a state of fixed $I$ becomes
\begin{equation}
{V}_{\rm eff}^I(r^2) =  \lambda \left ( \frac{\lambda +
\frac{I^2}{\lambda}}{2 \, m \, r^2} + \frac{m \, \omega^2  \,
r^2}{2 \lambda} \right ) \, .
\end{equation}
and the radial Schr\"odinger equation becomes:
 \begin{equation}
\left[ -\frac{1}{2 m } \frac{1}{r} \partial_r r \partial_r +
\lambda \left ( \frac{\lambda + \frac{I^2}{\lambda}}{2 \, m \,
r^2} + \frac{m \, \omega^2 \, r^2}{2 \lambda} \right ) \right]
\psi_I(r) = E_I \psi(r)  \,
 \end{equation}
This can be solved exactly.  The energies of the lowest-lying
state for any $I$ are given by
 \begin{eqnarray}
 E_I & = & \omega \left( 1+ \sqrt{\lambda^2 + I^2} \right )\nonumber \\
  & = & \omega \left ( \lambda + 1 + \frac{I^2}{2 \lambda}  + {\cal O}(\lambda^{-2}) \right
  ) \, .
 \label{EI} \end{eqnarray}
Note that these energies at large $\lambda$ accord with our large
$N_c$ expectations: the overall energy of the states go as
$\lambda \sim N_c$ as do the mass of baryons while the energy
splitting between states with different $I^2$ (of order 1) scales
as $\lambda^{-1} \sim N_c^{-1}$ as does, for example, the
N-$\Delta$ mass splitting.

Before putting a chemical potential on this system, it is useful
to show that a semiclassical approach analogous to that used from
Skyrmions reproduces the leading large $\lambda$ (large $N_c$
results) for both the overall energy and the splittings.  As a
first step one needs to find the analog of the soliton: a static
solution of the classical equations of motion.  This amounts to
finding $r_0$, the value of $r$ which minimizes the potential.  It
is straightforward to find $r_0$ and the value of the potential at
its minimum, $E_{0}$
\begin{eqnarray}
r_{0} & = &\sqrt{\frac{\lambda}{m \omega}} \nonumber \\
E_{0} & = & \lambda \omega
\end{eqnarray}
Note that $E_{0}$ is the analog of the ``static soliton mass''
$M_0$ appearing in e.g. Eq.(\ref{ll}); a
direct comparison with Eq.~(\ref{EI}) shows that it accurately
predicts the leading contribution to the energy.  There is
collective angular motion around the minimum. The corresponding
moment of inertia is $\Lambda_0 = \Lambda(r_0)$ with $\Lambda(r) = m r^2$.
It is easy to see that the requantization of the collective motion---in
analogy to the standard treatment of the Skyrme model---yields a
collective motion contribution to the energy of
 \begin{equation}
 E^{\rm coll}_I= \frac{I^2}{2 m r_0^2}= \frac{\omega I^2}{2
 { \lambda}}
 \end{equation}
which directly reproduces the leading $I$-dependent contribution
of Eq.~(\ref{EI}).  Thus the analog of the  standard Skyrmion
semiclassical approach correctly describes the dynamics of this
toy model up to the accuracy for which it is supposed to work.

In doing this comparison to Eq.~(\ref{EI}) the classical
``soliton'' correctly gives the order $\lambda \sim N_c$ part of
the exact energy.  However, there is also a term of order unity.
In the Skyrme model terms of order $N_c^0$  correspond to the zero
point fluctuations of the pion fields about the soliton background.
In Skyrme models they are difficult to compute and generally
neglected.  By analogy, in this model the $\lambda^0$ piece of the
energy should correspond to the zero point motion of the internal
degree of freedom.  This zero-point energy is given by 1/2 the
small amplitude vibration frequency about the minimum:
\begin{equation}
E^{\rm zero-point} =\frac{1}{2} \, \sqrt{\frac { \left . \frac{d^2
V}{d r^2 } \right |_{r_0} }{m}} =  \omega \; .
\end{equation}
It is worth noting that this exactly reproduces the term of order
$\lambda^0 \sim N_c^0$ in Eq.~(\ref{EI})

Now consider what happens if we add a chemical potential for $I$
to the Hamiltonian: $\hat{H}'=\hat{H}-\mu_I \hat{I}$.  The exact
eigenstates of $\hat{H}'$ corresponding to the lightest state with
fixed $I$ is given by $E_I'(\mui )= E_I - \mui I$.  Since we are
interested in the effect of the chemical potential, it is useful
to focus on the difference of the energy from the ground state
``baryon'' at $\mu_I=0$
 \begin{equation}
  \Delta E_I(\mui ) \equiv E_I'(\mui ) - E_{1/2} =
  \omega \left(\sqrt{\lambda^2 + I^2} - \sqrt{\lambda^2 + \frac{1}{4}} \right
  )- \mui I \; , \label{DeltaE}
  \end{equation}
which is plotted in Fig.~(\ref{Toyfig}) for a number of different
values of $I$ for the case $\lambda=15$.

{ We can now proceed to treat the problem semiclassically using an algorithm analogous
to the one in the previous section}. If one implements this assuming
$\mu_I \sim {\cal O}(1)$, one has a ``soliton'' whose { static}
energy, in analogy to Eq.(\ref{static}), is obtained by minimizing
\begin{equation}
V_{st}(r,\mui)=  V^{I=0}_{eff}(r) { -} \frac12 \ \Lambda(r) \  \mu_I^2 =
\lambda \left ( \frac{\lambda}{{ 2} m r^2} + \frac{ m \omega^2 r^2}{2 \lambda} \right )
-\frac12  m r^2 \mu_I^2
\; .
\end{equation}
Minimizing this with respect to $r$ to obtain the analog of the
soliton hegdehog mass $M_H(\mui)$ (see Eq.(\ref{hed})) one gets
\begin{eqnarray}
r^{(\mu_I)}_0 & = & \frac{\lambda^{1/2}}{m^{1/2}( \omega^2-\mui^2)^{1/4}} \nonumber \\
E_H(\mu_I) & = & V_{st}(r^{(\mu_I)}_0,\mui) = \lambda \sqrt{\omega^2 - \mui^2} \nonumber \\
 \Delta E_{H}(\mu_I)& =& E_{H}(\mu_I)- E_{H}(0)
 \label{SCsol}
 \end{eqnarray}
where $\Delta E_{H}(\mu_I)$ is the shift in the classical
``hedgehog mass'' due to the chemical potential.
 It is also
plotted in Fig.~(\ref{Toyfig}) for the case $\lambda=15$.

\begin{figure}
    \centering
        \subfigure[$\Delta E_I'$  versus chemical potential]{
            \label{Toy1}
            \includegraphics[width=3in]{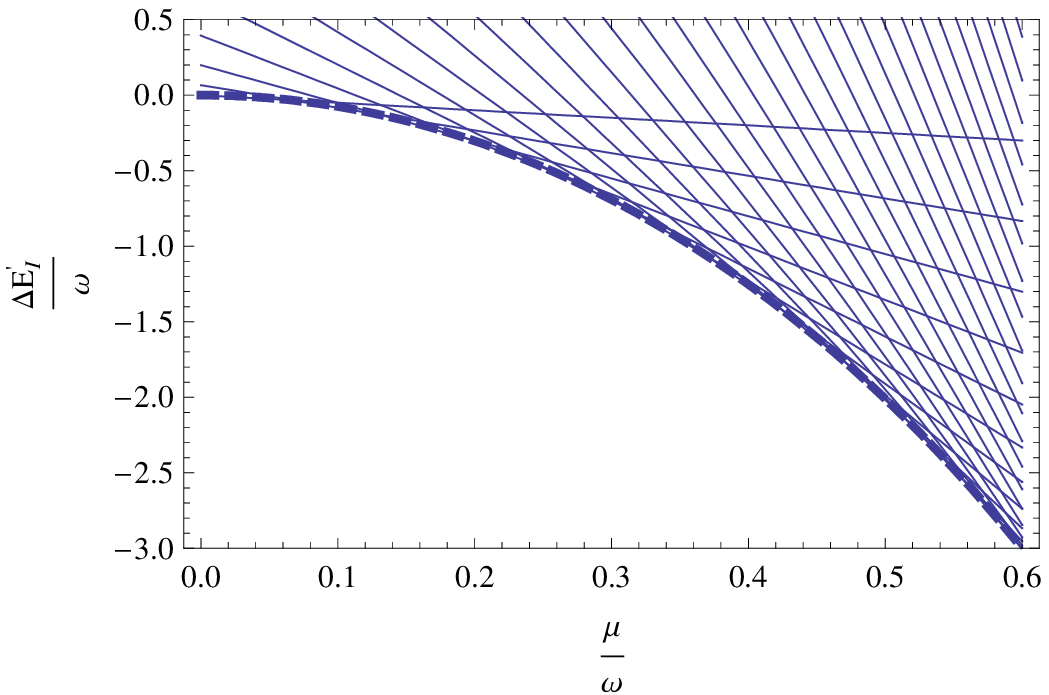}
        } \\
        \subfigure[The small $\mu_I$ regime]{
            \label{Toy2}
            \includegraphics[width=3in]{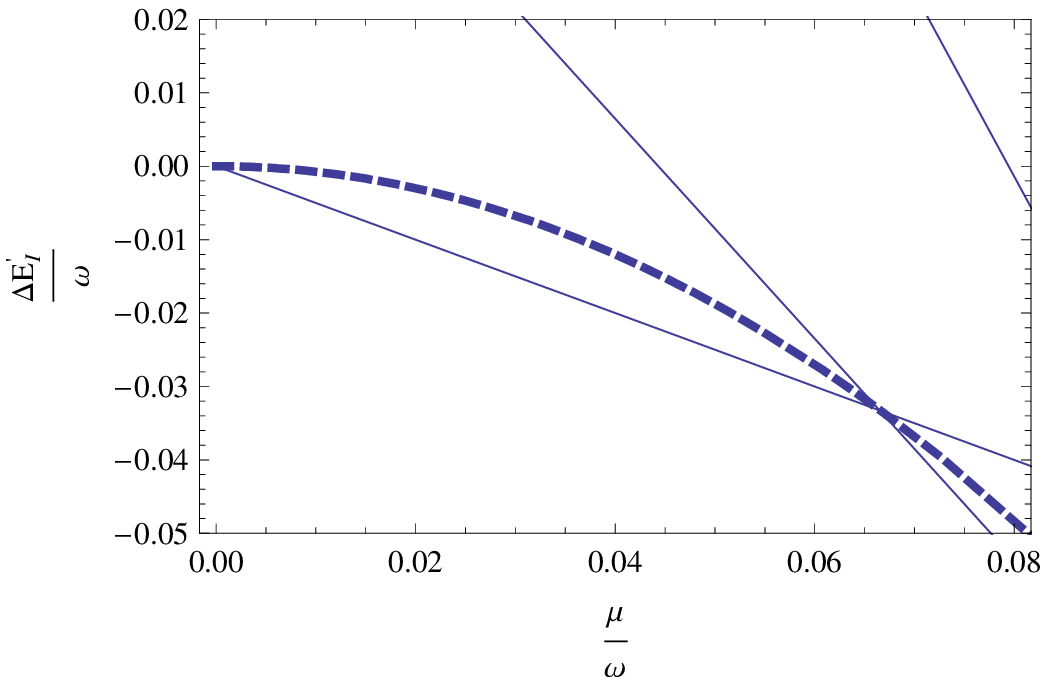}
        }
\caption{Energy shifts as defined in Eq.~(\ref{DeltaE}) (in units
of $\omega$) versus the chemical potential $\mu_I$ for
$\lambda=15$. The solid lines represent exact solutions.  At
$\mu_I=0$ the lowest line corresponds to $I=1/2$, the line above
it to $I=3/2$ with $I$ increasing by one unit each line up.  The
dashed line corresponds to the semiclassical solution of
Eq.~(\ref{SCsol}) Figure (a) goes out to $\mui=.6$ and illustrates
how the semiclassical result follows the minimum of the exact
solution. Figure (b) shows the low $\mu_I$ region and illustrates
the breakdown of the semiclassical treatment at very small
$\mu_I$.}
    \label{Toyfig}
    \end{figure}

It is apparent from these plots that the { classical} treatment
of the ``soliton'' works in the sense that {
it describes the shift in the
minimum energy solution as a function of $\mui$}. Moreover,
it is equally clear {\it how} it works.  Although each quantum
level is linear in $\mu_I$, as $\mu_I$ increases the quantum
levels cross and the value of $I$ of the lowest state increases.
The { classical} calculation tracks this level crossing---in a
continuous way.  Thus it is able to accurately reflect the
behavior of the minimum energy state when $\mu_I$ is large enough
so that $I \gg 1$.  Parametrically, this regime occurs for $\mu_I
/\omega \gg 1/\lambda \sim 1/N_c$ { where, therefore, rotational
corrections can be neglected}.  { Conversely} at small $\mu_I$ one
expects the semiclassical analysis to fail to reproduce the
shifts---and, { as it can be seen in Fig.2b}, this is indeed
the case.

Let us now look at the analog of the semiclassical calculation of
the proton's energy. { Following the same steps as in previous
section we get for the present toy model}
\begin{equation}
E_p^{sc} = E_0^{(\mui)} - \frac{\mui}{2} + \frac{1}{8 \Lambda^{(\mui)} }
\end{equation}
where
\begin{eqnarray}
E_0^{(\mui)} & = & V^{I=0}_{eff}(r_0^{(\mui)}) = E_H^{(\mu_I)} + \frac12 \Lambda^{(\mui)} \ \mui^2 \nonumber \\
\Lambda^{(\mui)} &=& m \ \left[ r_0^{(\mui)}\right]^2
\end{eqnarray}
The explicit form of ``proton'' energy resulting from this calculation is given by
\begin{eqnarray}
E_p^{\rm sc}(\mu_I ) &=& \omega \left[ \left ( \lambda \frac{ 1- \frac{\mu_I^2}{2 \omega^2}}{\sqrt{1 - \frac{\mu_I^2}{\omega^2}}}
+ \frac{1}{8\lambda} \sqrt{1 - \frac{\mu_I^2}{\omega^2}} \right ) -
\frac{\mu_I}{2 \omega} \right]
 \label{Ep} \end{eqnarray}
For ease of comparison we introduce the
shift in the semiclassically calculated energy from $\mu_I=0$
defined by $\Delta_p^{\rm sc} = E_p^{\rm sc}(\mu_I ) - E_p^{\rm sc}(0)$.
The corresponding exact result from Eq.~(\ref{DeltaE}) is $\Delta_p^{\rm exact}= -\mu_I/2$.
Both the exact and semiclassical result are plotted in Fig.~\ref{Toyfig2}.
\begin{figure}
    \centering
            \includegraphics[width=3in]{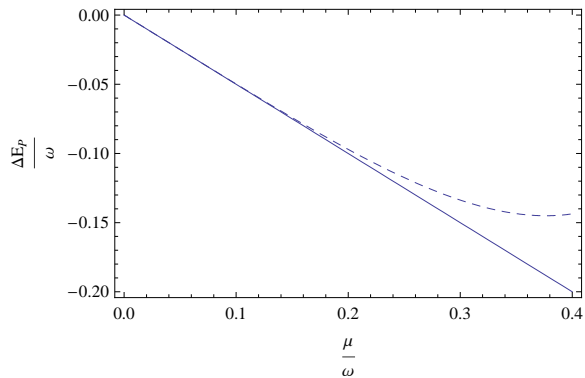}

\caption{Energy shifts for the ``proton''  ($I=1/2$) state as a
function of $\mu_I$ for the case of $\lambda=15$.  All energies are
given in units of $\omega$.  The solid line is the exact result;
the dashed line is the semiclassical result.}
    \label{Toyfig2}
    \end{figure}
It is apparent that the semiclassical procedure analogous to that
in the preceding section does {\it not} accurately reproduce the
exact result except at very small $\mu_I$ where nonlinearities are
negligible. The disease it suffers from is precisely the one identified
for the analogous Skyrmion calculation.  Namely, that in contrast
to the exact result, the calculated energy is nonlinear in $\mu_I$
with a nonzero curvature all the way down to $\mu_I=0$. It is also
clear {\it why} it fails. The ``soliton'' used in the calculation
was the minimum classical solution with fixed $\mu_I$.  Now it is
clear from Fig.~\ref{Toyfig}, that for $\mu_I$ of order one
($\lambda^0$) this classical configuration corresponds to quantum
states with relatively high $I$, {\it i.e.}, states which
parametrically have $I \sim \lambda$.  In situations where a
classical solution breaks a symmetry, and the underlying quantum
system is in the semiclassical limit, a classical configuration is
associated with a band of states which are all qualitatively
similar and all have energies which are near this minimum.  Thus,
when $\mu_I$ is parametrically of order unity, the classical
solution is valid only for describing states whose energy differs
from the minimum by energies of order $1/\lambda$.  The ``proton''
state is not in this class.

To summarize the results of a semiclassical treatment of this toy
model for $\mu_I \ne 0$: in the large $\lambda$ limit (analogous
to the large $N_c$ limit) and $\mu_I$ of order unity, the analog
of the classical soliton gives an accurate description of the
lowest-lying ``baryon'' states in terms of free energy---states
which have $I \sim \lambda$---but does not accurately describe the
``nucleon'' state.

\section{Discussion}

The toy model of the previous section illustrates the resolution
of the paradox of the existence of small
nonlinearities in the nucleon's energy as a
function of $\mu_I$ in Skyrme model calculations.  The key point
is that at large $N_c$ and zero temperature with $\mu_I \sim
N_c^0$ but below the phase transition ({\it i.e.}, $|\mu_I| <
m_\pi$), semiclassical calculations are formally valid for the
lowest-lying baryons states (which have $I \sim N_c$) but not for the nucleon.

This formal resolution of the paradox is useful for understanding
what is happening from a mathematical perspective.  However,
important phenomenological problems remain.  The crux of the issue
concerns the fact that in nature $N_c=3$ and the parameters of the
model are fit from the $N_c=3$ world. The high isospin states
critical to the formal resolution of this problem do not exist as
baryon resonances in the physical world.   Thus, there appears to
be no useful regime for which the semiclassical method for
describing how baryon free energies vary with $\mu_I$. The issue
is further complicated by the fact that $m_\pi$, while formally of
order $N_c^0$, is, in fact, very small due to approximate chiral
symmetry. Indeed, $m_\pi$ is smaller than $M_\Delta-M_N$ which is
formally of order $1/N_c$.  Thus if one were to increase $\mu_I$
starting from zero, even before the first level crossing occurs in
the baryons (the $I_3 = 3/2$ state of the $\Delta$ crossing the
proton), a phase transition occurs and the semiclassical methods
discussed here need to be modified.  It is not immediately obvious
how to do this.

Apart from phenomenology, an interesting formal question remains:
If the semiclassical method fails, how {\it does} one accurately
describe the nucleon (and other low isospin states, such as the
$\Delta$) for $\mu_I \neq 0$ in the large $N_c$ limit of the Skyrme
model)?  In the regime $| \mu_I | < m_\pi$, where the paradox
discussed here is manifest, this is easily accomplished. One first
does semiclassical quantization at $\mu_I=0$ and identifies the
quantum states.  Subsequently one imposes the chemical potential
at the quantum level on these states. Such a prescription is
guaranteed to give the correct linear behavior. However, the
problem gets more interesting when $| \mu_I | > m_\pi$.  In this
regime pions condense.  It is clear that new methods need to be
developed to describe baryons in this regime.

One of the authors (TDC) gratefully acknowledges the support of
the U. S. Department of Energy under grant no. DEFG02-93ER-40762.
NNS  acknowledges the support of CONICET (Argentina)
grant  PIP 6084,  and ANPCyT  (Argentina) grant  PICT04 03-25374 (NNS)

\end{document}